\documentclass[letterpaper]{article} 
\usepackage{aaai2026}  
\usepackage{times}  
\usepackage{helvet}  
\usepackage{courier}  
\usepackage[hyphens]{url}  
\usepackage{graphicx} 
\usepackage{natbib}  
\usepackage{caption} 
\usepackage{algorithm}
\usepackage{algpseudocode}

\usepackage{newfloat}
\usepackage{listings}
\DeclareCaptionStyle{ruled}{labelfont=normalfont,labelsep=colon,strut=off} 
\floatstyle{ruled}
\newfloat{listing}{tb}{lst}{}
\floatname{listing}{Listing}
\title{Perturbing Best Responses in Zero-Sum Games}
\author{
    Adam Dziwoki,
    Rostislav Hor{\v c}{\'i}k
}
\affiliations{
    Czech Technical University in Prague\\

    dziwoada@fel.cvut.cz, xhorcik@fel.cvut.cz
}

\usepackage{amssymb}
\usepackage{amsthm}
\usepackage{amsmath}
\usepackage{bm}
\usepackage{xfrac}
\usepackage{todonotes}

\DeclareMathOperator*{\argmax}{argmax}
\DeclareMathOperator*{\argmin}{argmin}

\theoremstyle{plain}
\newtheorem{theorem}{Theorem}
\newtheorem{corollary}[theorem]{Corollary}
\newtheorem{lemma}[theorem]{Lemma}

\theoremstyle{definition}

\newtheorem{example}{Example}

\newcommand{\tuple}[1]{\langle#1\rangle}

\newcommand{\uniform}{\mathcal{U}}
\newcommand{\br}{\mathsf{BR}}
\newcommand{\brv}{\mathsf{BRVal}}
\newcommand{\pbr}{\mathsf{\widetilde{BR}}}

\newcommand{\sm}{\mathsf{softmax}}
\newcommand{\resmat}[3]{\bm{#1}[#2,#3]} 

\begin{document}

\nocopyright
\maketitle

\begin{abstract}
This paper investigates the impact of perturbations on the best-response-based algorithms approximating Nash equilibria in zero-sum games, namely Double Oracle and Fictitious Play. More precisely, we assume that the oracle computing the best responses perturbs the utilities before selecting the best response. We show that using such an oracle reduces the number of iterations for both algorithms. For some cases, suitable perturbations ensure the expected number of iterations is logarithmic. Although the utility perturbation is computationally demanding as it requires iterating through all pure strategies, we demonstrate that one can efficiently perturb the utilities in games where pure strategies have further inner structure.
\end{abstract}

\begin{links}
    \link{Code}{https://github.com/geoborek/perturbing-best-responses}
\end{links}

\section{Introduction}

Computing Nash equilibria (NE) in two-player zero-sum games with huge strategy spaces is a computationally demanding problem.
Among algorithms approximating NE in such games, a substantial role is played by those based on \emph{best-response oracles} (BROs)
due to their ability to consider only a subspace of the strategy spaces. Prominent examples of such algorithms are 
\emph{Fictitious Play} (FP) introduced in~\cite{brown:fp-51} and \emph{Double Oracle} (DO)~\cite{mcmahan:etal:icml-03}.
The latter served as a basis for algorithms leveraging deep reinforcement learning to approximate best responses, 
such as \emph{Policy Space Response Oracles}~\cite{lanctot:etal:nips-17,mcaleer:arxiv-22,bighashdel:etal:ijcai-24}.

It is known that each two-player zero-sum game has an approximated $\varepsilon$-NE of logarithmic size in 
the number of pure strategies $n$ for given $\varepsilon>0$~\cite{althofer:laa-94,lipton:young:stoc-94}.
On the other hand, it is not difficult to construct games where FP and DO need at least $n$ iterations 
to find an $\varepsilon$-NE. This raises a natural question whether there is a BRO-based 
algorithm computing $\varepsilon$-NE in a logarithmic number of iterations. \cite{hazan:koren:stoc-16}
answered this question negatively as they proved that any (randomized) BRO-based algorithm needs 
at least $\Omega(\sqrt{n}/\log^3 n)$ iterations. At the same time, they introduced quite a complicated  
BRO-based algorithm providing a quadratic speed up $O(\sqrt{n}/\varepsilon^2)$ (up to a poly-logarithmic factor). 
Consequently, it follows that one needs to make BRO more powerful to achieve logarithmic 
number of iterations.

One way to make BRO stronger is by adding perturbations. Such an oracle perturbs the utilities before 
computing a best response for a given mixed strategy. We call such an oracle \emph{perturbed BRO} (PBRO).
A PBRO-variant of FP known as \emph{Stochastic Fictitious Play} (SFP) was introduced in~\cite{fundenberg:kreps:geb-93}.
Its convergence to NE in zero-sum games was proven in~\cite{hofbauer:sandholm:econometrica-02}.
In this paper, we prove that SFP achieves the logarithmic complexity in the number of pure strategies $n$
in expectation (note that PBRO makes SFP a randomized algorithm).
Analogously, we define a PBRO-based variant of DO called \emph{Stochastic Double Oracle} (SDO). Although 
its complexity in terms of $n$ is still an open problem, we prove that perturbations ensure logarithmic behavior
in expectation for some examples where DO needs $O(n)$ iterations, namely for examples introduced 
in~\cite{zhang:sandholm:ijcai-24}. We also tested experimentally on other games that perturbations reduce 
the number of iterations. However, it turned out that the perturbations do not accelerate the convergence in random games.
Implementing perturbations into a BRO for normal-form games requires iterating through all pure strategies 
to perturb the utility, which is not computationally efficient. However, for games whose strategy spaces 
have an inner structure like \emph{partially-observable stochastic game} (POSG), one can efficiently perturb only the rewards 
for transitions or terminal states. Even though such perturbations do not precisely correspond to the perturbations
in normal-form games, we experimentally show that they are able to reduce the number of iterations. 
More precisely, we demonstrate that on POSGs from~\cite{zhang:sandholm:ijcai-24} and a path-planning game 
where one player looks for the shortest path in a grid, while the other player might choose any edge and multiply its 
cost by a fixed coefficient.

\section{Related Work}

The convergence of FP was studied in several papers focusing mainly on the convergence rate w.r.t. $\varepsilon$
instead of the size of the game. FP converges to NE in zero-sum games, as proved in~\cite{robinson-51}. 
Its convergence rate is one of the oldest open problems in game theory, known as Karlin's conjecture~\cite{karlin-59}.
However, there are several partial results~\cite{abernethy:etal:soda-21,daskalakis:pan:focs-14}.
Recently, \emph{Anticipatory Fictitious Play} (AFP) was introduced~\cite{cloud:etal:ijcai-23}. AFP still needs 
$O(n)$ iterations in the worst case, but provides often faster convergence. Moreover, it calls the BRO four times in 
an iteration, whereas FP only calls the BRO two times. It is known that perturbations are related to the 
regularization~\cite{hofbauer:sandholm:econometrica-02,abernethy:etal:colt-14}. 
A regularized versions of AFP (PU and OMWU) were investigated in~\cite{cen:etal:jmlr-24}. 
Both variants need at most $O(\log n)$ iterations,
but they maintain probability distributions over all pure strategies, making them unsuitable for large games.

It is easy to see that DO needs $\Theta(n)$ iterations. However, to our knowledge, there is no deeper 
theoretical study of its convergence rate. Exponential lower bounds for DO applied to POSGs were obtained 
in~\cite{zhang:sandholm:ijcai-24}. Further variants of DO were investigated in~\cite{mcaleer:arxiv-22} but without any 
convergence guarantees.

\section{Background}

This section introduces our notation and the necessary background for the paper.
The set of natural numbers $\{1,\ldots,n\}$ is denoted $[n]$. The real-valued vectors are denoted by bold lowercase letters, e.g.
$\bm{p}\in\mathbb{R}^n$. Its $i$-th component is denoted $p_i$. The projection into the $i$-th component 
is the map $\pi_i(\bm{p})=p_i$. The vectors from the standard basis of $\mathbb{R}^n$ 
are denoted $\bm{e}_1,\ldots,\bm{e}_n$. Thus $\pi_i(\bm{e}_j)=1$ iff $i=j$ and $\pi_i(\bm{e}_j)=0$ otherwise.
Similarly, bold uppercase letters denote matrices, e.g. $\bm{M}\in\mathbb{R}^{m\times n}$. The entry of $\bm{M}$ 
in a row $i\in[m]$ and a column $j\in[n]$ is denoted $\bm{M}(i,j)=\bm{e}_i^\top\bm{M}\bm{e}_j$. 
The symbol $\Delta_n$ stands for the simplex of probability distributions over $[n]$. 
We identified the members of $\Delta_n$ with vectors $\bm{p}\in[0,1]^n$ such that $\sum_{i=1}^m p_i=1$.
Given sets of indexes $R\subseteq[m]$ and $C\subseteq[n]$, $\resmat{M}{R}{C}$ is the submatrix of $\bm{M}$ having
only rows with indexes in $R$ and columns with indexes in $C$.

This paper focuses on 2-player finite zero-sum games that can be identified with matrix games if we index the sets of
pure strategies by natural numbers.  
A \emph{matrix game} is given by a matrix $\bm{M}\in\mathbb{R}^{m\times n}$. The row indexes $[m]$ correspond to the 
set of pure strategies of the \emph{row player}. Analogously, the column indexes $[n]$ correspond to the pure strategies 
of the \emph{column player}. The entry $\bm{M}(i,j)$ denotes the game's outcome if the row player chooses 
the $i$-th strategy and the column player the $j$-th strategy. 
A \emph{mixed strategy} for the row player is a probability distribution $\bm{p}\in\Delta_m$ and analogously for the column player.
The expected outcome of the game when the row (resp. column) player plays $\bm{p}\in\Delta_m$ (resp. $\bm{q}\in\Delta_n$) 
can be expressed as $\bm{M}(\bm{p},\bm{q})=\bm{p}^\top\cdot\bm{M}\cdot\bm{q}$.
We assume that the entries in $\bm{M}$ represent losses (resp. rewards) for the row (resp. column) player. 
Thus, the row player chooses her strategy to minimize the expected outcome, whereas the column player wants to maximize it.  

The FP and DO algorithms rely on a best-response oracle for each player. 
Given a mixed strategy $\bm{q}\in\Delta_n$, the oracle for the row player computes the index of the best response 
and the corresponding value, namely
{\footnotesize
\begin{align*}
    \br_r(\bm{q})&\in\argmin_{i\in[m]} \pi_i(\bm{M}\bm{q}),&
    \brv_r(\bm{q})&=\min_{i\in [m]} \pi_i(\bm{M}\bm{q})
\end{align*}}
Analogously for $\bm{p}\in\Delta_m$, the column player's oracle computes
{\footnotesize
\begin{align*}
    \br_c(\bm{p})&\in\argmax_{j\in[n]} \pi_j(\bm{p}^\top\bm{M}),&
    \brv_c(\bm{p})&=\max_{j\in [n]} \pi_j(\bm{p}^\top\bm{M})
\end{align*}}
Note that $\br_r(\bm{q})$ and $\br_c(\bm{p})$ are pure strategies. In general, there might be several best responses 
for a given mixed strategy, and any of them can be returned by the oracle.

Given $\varepsilon\geq 0$, a pair of strategies $\tuple{\bm{p}^*,\bm{q}^*}$ is said to be 
\emph{$\varepsilon$-Nash equilibrium} ($\varepsilon$-NE) of the game $\bm{M}$ if the following inequalities hold:
{\footnotesize
\begin{align*}
   \bm{M}(\bm{p}^*,\bm{q}^*)-\brv_r(\bm{q}^*)\leq\varepsilon,
   \brv_c(\bm{p}^*)-\bm{M}(\bm{p}^*,\bm{q}^*)\leq\varepsilon
\end{align*}}
A \emph{Nash equilibrium} (NE) of the game $\bm{M}$ is $0$-NE.
Given a pair of strategies $\tuple{\bm{p},\bm{q}}$, a sufficient condition for the pair to form $\varepsilon$-NE is 
\begin{equation}\label{eq:termination}
\brv_c(\bm{p})-\brv_r(\bm{q})\leq\varepsilon
\end{equation}
that we use as a termination condition for FP and DO.

This paper investigates the influence of perturbations before selecting the best response. 
Let $\bm{u}$ and $\bm{v}$ be random vectors with respective dimensions $m,n$ whose 
components are i.i.d. random variables coming from a given distribution. \emph{Perturbed best responses}
are defined as follows:
\begin{align*}
    \pbr_r(\bm{q})\in\argmin_{i\in [m]} \pi_i(\bm{M}\bm{q}-\bm{u}),\\
    \pbr_c(\bm{p})\in\argmax_{j\in [n]} \pi_j(\bm{p}^\top\bm{M}+\bm{v})
\end{align*}
When we use the perturbed best responses in FP or DO, we assume
that the probability distribution for the perturbations vectors $\bm{u},\bm{v}$ is fixed.

This paper considers two common distributions \emph{uniform} $\mathcal{U}(a,b)$ on the interval $[a,b]$
and \emph{Gumbel} $G(\mu, \beta)$.
The Gumbel distribution $G(\mu, \beta)$ 
is a continuous probability distribution with a CDF given by
$F(x) = e^{-e^{-(x-\mu)/\beta}}$
where $\mu$ is a location and $\beta > 0$ a scale.
The Gumbel distribution is tightly related to the well-known function 
$\sm$, see  e.g.~\cite{goodfellow:etal:16}, mapping $\bm{x}\in\mathbb{R}^n$ to a probability distribution 
$\sm(\bm{x})\in\Delta_n$
defined by
\[\pi_i(\sm(\bm{x}))=\frac{e^{x_i}}{\sum_{j=1}^n e^{x_j}}\]

The following lemma is due to~\cite{gumbel:54} and is called \emph{Gumbel-max trick} in the machine learning 
community; for details see~\cite[Chapter~3]{train:09} or \cite{franke:degen:23}. Given $\bm{x}\in\mathbb{R}^n$,
the Gumbel-max trick allows to sample from $\sm(\bm{x})$ by taking $\argmax$ of the perturbed values $x_i$.

\begin{lemma}\label{l:gumbel}
Let $\bm{x}\in\mathbb{R}^n$ and $i^*\in\{1,\ldots,n\}$. 
Let $\bm{z}\in\mathbb{R}^n$ be a random vector whose components $z_i$ be drawn i.i.d. from $G(0,\beta)$. 
Then the probability
\[P(\argmax_{i\in[n]} \pi_i(\bm{x}+\bm{z}) = i^*) = \pi_{i^*}(\sm(\bm{x}/\beta))\]
\end{lemma}
Dually, it follows from Lemma~\ref{l:gumbel} that
\[P(\argmin_{i\in[n]} \pi_i(\bm{x}-\bm{z}) = i^*) = \pi_{i^*}(\sm(-\bm{x}/\beta))\]
as $\argmin_i \pi_i(\bm{x}-\bm{z}) = i^*$ iff $\argmax_i \pi_i(-\bm{x}+\bm{z}) = i^*$. 
Consequently, if we perturb the utilities with the Gumbel distribution $G(0,\beta)$, 
the perturbed best-response $\pbr_r(\bm{q})$ is sampled from 
$\sm(\bm{M}\bm{q}/\beta)$ and analogously $\pbr_c(\bm{p})$ from $\sm(-\bm{p}^\top\bm{M}/\beta)$.

To prove our result on SFP, we need to recall the \emph{randomized exponentially weighted forecaster} (REWF),
introduced in~\cite[Chapter~4]{cesa-bianchi:lugosi:plg-06}. 
Let $\bm{M}\in [0,1]^{m\times n}$ be a $m\times n$-matrix viewed as a loss function $[m]\times[n]\to[0,1]$. 
Consider a ``sort of'' game between 
a player and an opponent with $T\geq 1$ rounds.
In the round $t$, the player chooses an action $i_t\in[m]$ and the opponent an action $j_t\in[n]$,
making the player to suffer the loss $\bm{M}(i_t,j_t)$. The player's goal is to minimize the cumulative loss
$\sum_{t=1}^T \bm{M}(i_t,j_t)$.
REWF is a randomized strategy for the player based on the opponent's previously selected actions. 
In the round $t$, REWF samples $i_t$ based on $j_1,\ldots,j_{t-1}$ from the distribution
$\sm(-\eta\sum_{s=1}^{t-1} \bm{M}\bm{e}_{j_s})$, where $\eta>0$ is a parameter. 
Following REWF ensures an upper bound 
on the regret, i.e., the difference between the cumulative loss and 
the best fixed player's action $i$; it immediately follows from \cite[Theorem~2.2 and Lemma~4.1]{cesa-bianchi:lugosi:plg-06}.
\begin{corollary}\label{c:rewf}
    Let $\bm{M}\in[0,1]^{m\times n}$, $T\geq 1$, $\eta>0$, and $\delta\in(0,1)$. REWF satisfies, 
    with probability $1-\delta$,
    {\footnotesize\[\sum_{t=1}^T \bm{M}(i_t,j_t)-\min_{i\in[m]}\sum_{t=1}^T \bm{M}(i,j_t)\leq 
    \frac{\ln m}{\eta}+\frac{T\eta}{8}+\sqrt{\frac{T}{2}\ln\frac{1}{\delta}}\]}
\end{corollary}

\begin{algorithm}[t]
\caption{Stochastic Fictitious Play}\label{alg:sfp}
    \begin{algorithmic}
        \Require Initial strategies $k,l$; $\varepsilon>0$; a probability distribution for $\pbr_r$ and $\pbr_c$
        \Ensure $\varepsilon$-NE $\tuple{\bm{p}^*, \bm{q}^*}$
        \State $t \gets 1$
        \State $\bm{p} \gets \bm{e}_k$, $\bm{q} \gets \bm{e}_l$
        \State $lb \gets \brv_r(\bm{q})$, $ub \gets \brv_c(\bm{p})$ 
        \While {$ub-lb > t\varepsilon$}
            \State $t \gets t + 1$
            \State $i \gets \pbr_r(\bm{q})$, $j \gets \pbr_c(\bm{p})$
            \State $\bm{p} \gets \bm{p} + \bm{e}_{i}$, $\bm{q} \gets \bm{q} + \bm{e}_{j}$
            \State $lb \gets \brv_r(\bm{q})$, $ub \gets \brv_c(\bm{p})$ 
        \EndWhile
        \State \Return $\tuple{\bm{p}^*,\bm{q}^*}=\tuple{\bm{p}/t,\bm{q}/t}$
    \end{algorithmic}
\end{algorithm}

\section{Stochastic Fictitious Play}

\emph{Stochastic Fictitious Play} (SFP) is a randomized version of \emph{Fictitious Play} (FP). 
Its pseudocode is shown in Algorithm~\ref{alg:sfp}. We expand it with the termination 
condition~(\ref{eq:termination}) so that it stops after reaching $\varepsilon$-NE. 
Note that the best-response values $\brv_r(\bm{q})$, $\brv_c(\bm{p})$ are not perturbed, 
hence the termination condition is not affected by the perturbations.
The algorithm maintains two vectors 
$\bm{p}$ and $\bm{q}$ representing multisets of already played pure strategies. In each iteration, 
SFP finds perturbed best responses 
w.r.t. $\bm{p}$ and $\bm{q}$ respectively\footnote{Formally, $\bm{p}$, $\bm{q}$ are not 
probability distributions. We assume w.l.o.g. that the best-response oracle works for them as well.}. 
When the termination condition
is satisfied, SFP returns the average of all the 
played strategies so that the resulting vectors $\bm{p}^*,\bm{q}^*$ form probability distributions.
Note that even though Algorithm~\ref{alg:sfp} operates with vectors $\bm{e}_i\in\mathbb{R}^m$, $\bm{e}_j\in\mathbb{R}^n$, 
SFP can be implemented efficiently without considering all the dimensions by storing only the multisets of 
pure strategies played by the players up to the current iteration.

Using the Gumbel-max trick (see Lemma~\ref{l:gumbel}), 
we show that SFP combined with Gumbel perturbations corresponds to the computation 
when both players apply the randomized exponentially weighted forecaster (REWF) against each other.
W.l.o.g. we assume that the matrix game $\bm{M}\in\mathbb{R}^{m\times n}$ is normalized into $[0,1]$
and $m\leq n$.

\begin{theorem}\label{th:SFP}
    Let $\bm{M}\in[0,1]^{m\times n}$ be a matrix game such that $m\leq n$ and $\varepsilon>0$. 
    An $\varepsilon$-NE can be computed by SFP with perturbations sampled from $G(0,\beta)$ 
    in $O(\frac{\log n}{\varepsilon^2})$ expected iterations, where 
    $\beta=\frac{2+\sqrt{2\ln n}}{\varepsilon\sqrt{8\ln n}}$. 
\end{theorem}

\begin{proof}
    We first prove that SFP finds $\varepsilon$-NE in $O(\frac{\log n}{\varepsilon^2})$ iterations with probability 
    at least $1/2$. We let SFP to iterate for $T=\left(\frac{2+\sqrt{2\ln n}}{\varepsilon}\right)^2\in O(\frac{\log n}{\varepsilon^2})$ iterations.

    Let $\bm{p}_t$, $\bm{q}_t$ denote the vectors $\bm{p},\bm{q}$ at the iteration $t\leq T$.
    Further, let $i_t=\pbr_r(\bm{q}_{t-1})$ and $j_t=\pbr_c(\bm{p}_{t-1})$. 
    At the iteration $t$, we have $\bm{q}_{t-1}=\sum_{s=1}^{t-1}\bm{e}_{j_s}$.
    Thus $\bm{M}\bm{q}_{t-1}=\sum_{s=1}^{t-1} \bm{M}\bm{e}_{j_s}$.
    By Lemma~\ref{l:gumbel}, $i_t$ is sampled
    from the distribution $\sm(-\eta\sum_{s=1}^{t-1}\bm{M}\bm{e}_{j_s})$ for $\eta=1/\beta$. 
    By our choice of $\beta$ and $T$, we have $\eta=\sqrt{8\ln n/T}$.
    Consequently, Corollary~\ref{c:rewf} for $\delta=1/4$ implies with probability at least $3/4$, using also $m\leq n$ and $\ln 4\leq 2$,
    {\footnotesize
    \begin{align*}
    \sum_{t=1}^T \bm{M}(i_t,j_t)-\min_{i\in[m]}\sum_{t=1}^T \bm{M}(i,j_t)&\leq 
    \frac{\ln m}{\eta}+\frac{T\eta}{8}+\sqrt{\frac{T}{2}\ln 4}\\
    \leq \frac{\ln n}{\eta}+\frac{T\eta}{8}+\sqrt{T}&=\sqrt{T}\left(1+\sqrt{\frac{\ln n}{2}}\right)
    \end{align*}}
    
    Applying Corollary~\ref{c:rewf} to the matrix $\bm{1}-\bm{M}^\top$, 
    we can derive an analogous bound for the column player with probability at least $3/4$
    {\footnotesize
    \[\max_{j\in[n]}\sum_{t=1}^T \bm{M}(i_t,j)-\sum_{t=1}^T \bm{M}(i_t,j_t)\leq 
    \sqrt{T}\left(1+\sqrt{\frac{\ln n}{2}}\right)
    \]}

    Consequently, both inequalities hold simultaneously with probability at least $1/2$ using 
    the well-known lower bound $P(A\cap B)\geq P(A)+P(B)-1$ on the probability of the intersection of two events 
    $A,B$. 
    Summing the above inequalities, we get
    {\footnotesize
    \begin{multline*}
        \max_{j\in[n]}\sum_{t=1}^T \bm{M}(i_t,j)-\min_{i\in[n]}\sum_{t=1}^T \bm{M}(i,j_t)
        \leq \sqrt{T}(2+\sqrt{2\ln n})
    \end{multline*}}
    SFP returns the pair $\tuple{\frac{1}{T}\sum_{t=1}^T \bm{e}_{i_t},\frac{1}{T}\sum_{t=1}^T \bm{e}_{j_t}}$.
    Note that {\footnotesize
    \begin{align*}
        ub=\brv_c\left(\frac{1}{T}\sum_{t=1}^T \bm{e}_{i_t}\right)&= 
        \max_{j\in [n]} \pi_j\left(\frac{1}{T}\sum_{t=1}^T \bm{e}_{i_t}^\top\bm{M}\right)\\
        &=\frac{1}{T}\max_{j\in [n]} \sum_{t=1}^T \bm{M}(i_t,j)
    \end{align*}}
    Analogously, $lb=\frac{1}{T}\min_{i\in [m]} \sum_{t=1}^T \bm{M}(i,j_t)$.
    Combining the above facts, we get
    $ub-lb\leq(2+\sqrt{2\ln n})/\sqrt{T}=\varepsilon$.
    Thus $\tuple{\frac{1}{T}\sum_{t=1}^T \bm{e}_{i_t},\frac{1}{T}\sum_{t=1}^T \bm{e}_{j_t}}$ forms $\varepsilon$-NE
    with probability at least $1/2$.

    To obtain an algorithm with a logarithmic expected number of iterations, we execute
    SFP for $T$ many iterations until we find $\varepsilon$-NE. It occurs in the first or second run in expectation.
\end{proof}

\begin{algorithm}[t]
\caption{Stochastic Double Oracle}\label{alg:sdo}
    \begin{algorithmic}
        \Require Initial strategies $k,l$; $\varepsilon>0$; a probability distribution for $\pbr_r$ and $\pbr_c$
        \Ensure $\varepsilon$-NE $\tuple{\bm{p}^*, \bm{q}^*}$
        \State $t \gets 1$
        \State $R \gets \{k\}$, $C \gets \{l\}$
        \State $\bm{p} \gets \bm{e}_k$, $\bm{q} \gets \bm{e}_l$
        \State $lb \gets \brv_r(\bm{q})$, $ub \gets \brv_c(\bm{p})$ 
        \While {$ub-lb > \varepsilon$}
            \State $t \gets t + 1$
            \State $\tuple{\bm{p},\bm{q}} \gets \mathrm{getNash}(\resmat{M}{R}{C})$
            \State $i \gets \pbr_r(\bm{q})$, $R \gets R\cup \{i\}$
            \State $j \gets \pbr_c(\bm{p})$, $C \gets C\cup \{j\}$
            \State $lb \gets \brv_r(\bm{q})$, $ub \gets \brv_c(\bm{p})$ 
        \EndWhile
        \State \Return $\tuple{\bm{p}^*,\bm{q}^*}=\tuple{\bm{p},\bm{q}}$
    \end{algorithmic}
\end{algorithm}

\section{Stochastic Double Oracle}

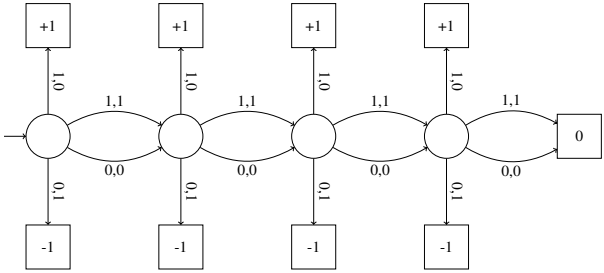
\begin{figure}[t]
    \centering
    \resizebox{.45\textwidth}{!}{
\begin{tikzpicture}[
    every node/.style={draw, minimum size=1cm},
    state/.style={circle},
    reward/.style={rectangle},
    node distance=2cm and 2cm,
  ]
  \node[state] (s) at (1,0) {};
  \node[state, right=of s] (a) {};
  \node[state, right=of a] (b) {};
  \node[state, right=of b] (c) {};
  \node[reward, right=of c] (d) {0};

  \draw[->] (0,0) to (s);
  
  \draw[->, bend right] (s) to node [draw = none, midway, below, yshift = 8pt] {0,0} (a);
  \draw[->, bend left] (s) to node [draw = none, midway, above, yshift=-8pt] {1,1} (a);

  \draw[->, bend right] (a) to node [draw = none, midway, below, yshift=8pt] {0,0} (b);
  \draw[->, bend left] (a) to node [draw = none, midway, above,yshift=-8pt] {1,1} (b);

  \draw[->, bend right] (b) to node [draw = none, midway, below,yshift=8pt] {0,0} (c);
  \draw[->, bend left] (b) to node [draw = none, midway, above,yshift=-8pt] {1,1} (c);
  
  \draw[->, bend right] (c) to node [draw = none, midway, below,yshift=8pt] {0,0} (d);
  \draw[->, bend left] (c) to node [draw = none, midway, above,yshift=-8pt] {1,1} (d);

  \node[reward, above=1.5 cm of s] (w1) {+1};
  \node[reward, below=1.5cm of s] (l1) {-1};
  \node[reward, above=1.5cm of a] (w2) {+1};
  \node[reward, below=1.5cm of a] (l2) {-1};
  \node[reward, above=1.5cm of b] (w3) {+1};
  \node[reward, below=1.5cm of b] (l3) {-1};
  \node[reward, above=1.5cm of c] (w4) {+1};
  \node[reward, below=1.5cm of c] (l4) {-1}; 

  \draw[->] (s) to node[draw = none, above, rotate=270,yshift =-8pt] {1,0} (w1);
  \draw[->] (s) to node [draw = none, above, rotate =270, yshift=-8pt] {0,1} (l1);
  \draw[->] (a) to node[draw = none, above, rotate=270,yshift =-8pt] {1,0} (w2);
  \draw[->] (a) to node[draw = none, above, rotate=270,yshift =-8pt] {0,1} (l2);
  \draw[->] (b) to node[draw = none, above, rotate=270,yshift =-8pt] {1,0} (w3);
  \draw[->] (b) to node[draw = none, above, rotate=270,yshift =-8pt] {0,1} (l3);
  \draw[->] (c) to node[draw = none, above, rotate=270,yshift =-8pt] {1,0} (w4);
  \draw[->] (c) to node[draw = none, above, rotate=270,yshift =-8pt] {0,1} (l4);
\end{tikzpicture}}
    \caption{An example of the 4-bit stochastic game.}
    \label{fig:non-unique}
\end{figure}

\begin{figure}[t]
    \centering
    \resizebox{.45\textwidth}{!}{
\begin{tikzpicture}[
    every node/.style={draw, minimum size=1cm},
    state/.style={circle},
    reward/.style={rectangle},
    node distance=2cm and 2cm
  ]

  \node[state] (s) at (1,0) {};
  \node[state, right=of s] (a) {};
  \node[state, right=of a] (b) {};
  \node[state, right=of b] (c) {};
  \node[reward, right=of c] (d) {0};
  \node[state, above=1.5cm of s] (e) {};
  \node[state, above=1.5cm of a] (f) {};
  \node[state, above=1.5cm of b] (g) {};
  \node[reward, above=1.5cm of c] (h) {+2};
  \node[state, below=1.5cm of s] (i) {};
  \node[state, below=1.5cm of a] (j) {};
  \node[state, below=1.5cm of b] (k) {};
  \node[reward, below=1.5cm of c] (l) {-2};
  \node[reward, above=1.5cm of e] (m) {+1};
  \node[reward, above=1.5cm of f] (n) {+1};
  \node[reward, above=1.5cm of g] (o) {+1};
  \node[reward, below=1.5cm of i] (p) {-1};
  \node[reward, below=1.5cm of j] (q) {-1};
  \node[reward, below=1.5cm of k] (r) {-1};

  \draw[->] (0,0) to (s);
  
  \draw[->, bend right] (s) to node [draw = none, below, yshift = 8pt] {0,0} (a);
  \draw[->, bend left] (s) to node [draw = none, above, yshift=-8pt] {1,1} (a);

  \draw[->, bend right] (a) to node [draw = none, below, yshift=8pt] {0,0} (b);
  \draw[->, bend left] (a) to node [draw = none, above,yshift=-8pt] {1,1} (b);

  \draw[->, bend right] (b) to node [draw = none, below,yshift=8pt] {0,0} (c);
  \draw[->, bend left] (b) to node [draw = none, above,yshift=-8pt] {1,1} (c);
  
  \draw[->, bend right] (c) to node [draw = none, below,yshift=8pt] {0,0} (d);
  \draw[->, bend left] (c) to node [draw = none, above,yshift=-8pt] {1,1} (d);

  \draw[->] (s) to node[draw = none, above, rotate=270,yshift =-8pt] {1,0} (e);
  \draw[->] (s) to node [draw = none, above, rotate =270, yshift=-8pt] {0,1} (i);
  \draw[->] (a) to node[draw = none, above, rotate=270,yshift =-8pt] {1,0} (f);
  \draw[->] (a) to node[draw = none, above, rotate=270,yshift =-8pt] {0,1} (j);
  \draw[->] (b) to node[draw = none, above, rotate=270,yshift =-8pt] {1,0} (g);
  \draw[->] (b) to node[draw = none, above, rotate=270,yshift =-8pt] {0,1} (k);
  \draw[->] (c) to node[draw = none, above, rotate=270,yshift =-8pt] {1,0} (h);
  \draw[->] (c) to node[draw = none, above, rotate=270,yshift =-8pt] {0,1} (l);

  \draw[->] (e) to node[draw = none, above, yshift = -8 pt] {0,1} (f);
  \draw[->] (f) to node[draw = none, above, yshift = -8 pt] {0,1} (g);
  \draw[->] (g) to node[draw = none, above, yshift = -8 pt] {0,1} (h);

  \draw[->] (i) to node[draw = none, above, yshift = -8 pt] {0,1} (j);
  \draw[->] (j) to node[draw = none, above, yshift = -8 pt] {0,1} (k);
  \draw[->] (k) to node[draw = none, above, yshift = -8 pt] {0,1} (l);

  \draw[->] (e) to node[draw = none, above, xshift = -8 pt, rotate = 270] {1,1} (m);
  \draw[->] (f) to node[draw = none, above, xshift = -8 pt, rotate = 270] {1,1} (n);
  \draw[->] (g) to node[draw = none, above, xshift = -8 pt, rotate = 270] {1,1} (o);

  \draw[->] (i) to node[draw = none, above, xshift = -8 pt, rotate = 270] {1,1} (p);
  \draw[->] (j) to node[draw = none, above, xshift = -8 pt, rotate = 270] {1,1} (q);
  \draw[->] (k) to node[draw = none, above, xshift =-8pt, rotate =270] {1,1} (r);

  \draw[->, bend right=45] (e) to node [draw = none, below,xshift=20pt, rotate = 270] {0,0} (m);
  \draw[->, bend left=45] (e) to node [draw = none, below ,xshift=-20 pt, rotate = 90] {1,0} (m);

    \draw[->, bend right=45] (f) to node [draw = none, below,xshift=20pt, rotate = 270] {0,0} (n);
  \draw[->, bend left=45] (f) to node [draw = none, below ,xshift=-20 pt, rotate = 90] {1,0} (n);

    \draw[->, bend right=45] (g) to node [draw = none, below,xshift=20pt, rotate = 270] {0,0} (o);
  \draw[->, bend left=45] (g) to node [draw = none, below ,xshift=-20 pt, rotate = 90] {1,0} (o);

  \draw[->, bend right=45] (i) to node [draw = none, above,xshift=8pt, rotate = 90] {0,0} (p);
  \draw[->, bend left=45] (i) to node [draw = none, below ,xshift=20 pt, rotate = 270] {1,0} (p);

  \draw[->, bend right=45] (j) to node [draw = none, above,xshift=8pt, rotate = 90] {0,0} (q);
  \draw[->, bend left=45] (j) to node [draw = none, below ,xshift=20 pt, rotate = 270] {1,0} (q);
\draw[->, bend right=45] (k) to node [draw = none, above,xshift=8pt, rotate = 90] {0,0} (r);
  \draw[->, bend left=45] (k) to node [draw = none, below ,xshift=20 pt, rotate = 270] {1,0} (r);
  
\end{tikzpicture}
    }
    \caption{An example of the 4-bit POSG.}
    \label{fig:unique}
\end{figure}
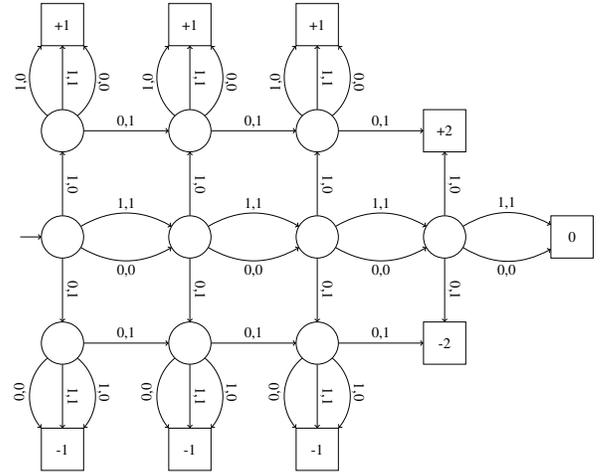

This section introduces \emph{Stochastic Double Oracle} (SDO), a randomized variant of Double Oracle (DO)
with perturbed best responses. Its pseudocode is shown in Algorithm~\ref{alg:sdo}. As DO, it maintains two sets of 
pure strategies $R,C$ for the row and column player, respectively. At each iteration, it computes NE for the 
induced subgame $\resmat{M}{R}{C}$. The termination condition is the same as in SFP.

In this paper, we take the first steps in investigating convergence for SDO. We start with two matrix games 
introduced in~\cite{zhang:sandholm:ijcai-24}
that are difficult to solve for DO, i.e., DO needs to iterate through all pure strategies to find $\varepsilon$-NE.  
We prove that SDO with suitable perturbations finds $\varepsilon$-NE in a logarithmic number of iterations in expectation for both.

\begin{example}\label{ex:non-unique}
    Let $n\geq 1$. In the first game, both players choose a number from $[n]$. The player who chooses the greater number wins.
    If the chosen numbers are equal, it is a draw. The corresponding square matrix $\bm{L}\in\mathbb{R}^{n\times n}$ is 
    defined as follows:
\[
\bm{L}=\left(
    \begin{array}{cccc}
        \phantom{-}0 & \phantom{-}1 & \cdots & 1\\
       -1 &  \phantom{-}0 & \cdots & 1\\
        \vdots & \phantom{-}\vdots & \ddots & \phantom{-}\vdots\\
       -1 &  -1 & \cdots & 0
    \end{array} 
\right)
\]
\end{example}
The game $\bm{L}$ has a unique pure $\varepsilon$-NE consisting of the last row and column
for any $0\leq\varepsilon<1$. Thus finding $\varepsilon$-NE for $\varepsilon<1$ is the same as finding NE. If DO is initialized with $k=l=1$ and 
the best responses are chosen adversarially (i.e., we always take the best response with the least index), DO needs $n$ iterations
to find the NE. Note that there are several candidates when DO looks for a best response to a row or column. 
On the other hand, when we apply SDO to $\bm{L}$, the perturbations allow us to select a candidate uniformly randomly. 

We prove our result on SDO by applying drift theory~\cite{kotzing:krejza:tcs-19}.
For technical reasons, we prove it w.l.o.g. for a dual game where the player who chooses the smaller number wins.
Its matrix $\bm{S}=\bm{L}^\top$ has a unique NE consisting of the first row and column. 

Let $k>1$. Consider the $k$-th row $\bm{x}$ of $\bm{S}$, i.e., $x_i=1$ for $i<k$, $x_k=0$, and $x_i=-1$ for $i>k$.
The following crucial lemma shows that $\pbr_c(\bm{e}_k)$ with uniform perturbations selects 
a best response uniformly among $\{1,\ldots,k-1\}$. Analogous lemma can be proven for the $k$-th column.
\begin{lemma}\label{l:non-unique}
    Let $\bm{x}=\tuple{1,\ldots,1,0,-1,\ldots,-1}\in\mathbb{R}^n$ be the $k$-th row of $\bm{S}$, 
    $k>1$, and 
    $\bm{z}=\tuple{z_1,\ldots,z_n}$ a random vector whose components 
    $z_i\sim\uniform(-1/2,1/2)$. 
    Let $I$ be the random variable $I=\argmax_{i\in[n]}\pi_i(\bm{x}+\bm{z})$.
    Then $E[I]=\frac{k}{2}$.
\end{lemma}
\begin{proof}
    Note that $x_i+z_i\sim\mathcal{U}(1/2,3/2)$ for $i<k$, $x_i+z_i\sim\mathcal{U}(-1/2,1/2)$ for $i=k$,
    and $x_i+z_i\sim\mathcal{U}(-3/2,-1/2)$ for $i>k$. Thus
    we have $E[I] = \sum_{i=1}^n i\cdot P(I=i)=\sum_{i=1}^{k-1} i\cdot P(I=i)$ because $P(I=i)=0$ for 
    $i\geq k$. Moreover, $P(I=i)=\frac{1}{k-1}$ is the uniform distribution on 
    $\{1,\ldots,k-1\}$ as the first $k-1$ components of $\bm{x}$ are equal.
    Consequently, $E(I)=\frac{1}{k-1}\sum_{i=1}^{k-1}i = \frac{k}{2}$.
\end{proof}

Next, we need to understand equilibria of the subgame $\resmat{S}{R}{C}$ in every iteration of SDO.
For notational simplicity, we index rows and columns in $\resmat{S}{R}{C}$ with the same indexes as 
in the original matrix $\bm{S}$. Further, we denote the probability simplices over $R,C$ by 
$\Delta_R,\Delta_C$, respectively.
\begin{lemma}\label{l:NE-cases}
    Let $R,C\subseteq [n]$ and $r = \min R$, $c = \min C$. 
    \begin{enumerate}
    \item If $r < c$, then $\tuple{\bm{e}_i,\bm{q}}$ is NE in $\resmat{S}{R}{C}$ for any $i\in\{k\in R\mid k<c\}$
    and $\bm{q}\in\Delta_C$. There are no other NEs. 
    \item Dually, if $c < r$, then $\tuple{\bm{p},\bm{e}_j}$ is NE in $\resmat{S}{R}{C}$ 
    for any $j\in\{k\in C\mid k<r\}$ and $\bm{p}\in\Delta_R$. There are no other NEs. 
    \item If $r=c$, then $\tuple{\bm{e}_r,\bm{e}_c}$ is the unique NE in $\resmat{S}{R}{C}$.
    \end{enumerate}
\end{lemma}
\begin{proof}
    For the first item, let $\bm{x}$ be an $i$-th row of $\bm{S}$ for $i\in\{k\in R\mid i<c\}$. 
    Note that $x_j=-1$ for all $j\in C$. Thus the $i$-th row in $\resmat{S}{R}{C}$ is equilibrial strategy for the 
    row player. A column player can play an arbitraty strategy.
    The second item is proven analogously as the first one.

    The last claim follows if we note that the $r$-th row in $\resmat{S}{R}{C}$ is of the form $\tuple{0,-1,\ldots,-1}$
    and $c$-th column in $\resmat{S}{R}{C}$ is of the form $\tuple{0,1,\ldots,1}$.
\end{proof}

\begin{theorem}\label{th:SDO-non-unique}
    SDO finds NE of $\bm{S}$ in $O(\log n)$ expected number of iterations. 
\end{theorem}
\begin{proof}
    We assume $k,l=n$, which is the worst possible start for SDO on $\bm{S}$. 
    Let $R_t, C_t$ be the row and column strategies in the iteration $t$. 
    Define $r_t=\min R_t$ and $c_t=\min C_t$.
    Next, we define $X_t=\max\{r_t,c_t\}$. Apparently, we have $X_{t+1}\leq X_t$ since $r_{t+1}\leq r_t$ and $c_{t+1}\leq c_t$.
    Once $X_t=1$, SDO will find the NE in the next iteration. 
    Thus, we want to show $E[T]\in O(\log n)$ where $T=\inf\{t\mid X_t\leq 1\}$.
    This can be proven by \cite[Corollary~17]{kotzing:krejza:tcs-19}, if we show that 
    $X_t-E[X_{t+1}\mid X_1,\ldots,X_t]\geq \delta X_t$ for some $\delta>0$.
    Regarding the sequence $X_1,X_2,\ldots$, we will consider only subsequence $X_1,X_3,\ldots$ 
    of odd iterations $t$ as SDO needs often two iterations to improve 
    strategies for both players. In other words, we will prove 
    $X_t-E[X_{t+2}\mid X_1,X_3\ldots,X_t]\geq \delta X_t$ for some $\delta>0$.
    To simplify the notation, we denote the conditional expectations without the condition in the rest 
    of proof. For instance, $E[X_{t+2}]$ means $E[X_{t+2}\mid X_1,X_3,\ldots,X_t]$.

    Let $t$ be an odd iteration. There are two cases. We prove the case if $r_t\leq c_t$. 
    The proof for the case when $r_t\geq c_t$ is analogous.
    If $r_t\leq c_t$, then $\tuple{\bm{e}_i,\bm{q}}$ is NE for $\resmat{S}{R_t}{C_t}$ 
    for some $i\leq c_t$ and some $\bm{q}\in\Delta_C$ by Lemma~\ref{l:NE-cases} (items 1 and 3). 
    Consequently, the best response 
    for the column player is $j=\pbr_c(\bm{e}_{i})<i\leq c_t$. Thus $j$ becomes the new 
    minimum of $C_{t+1}$, i.e., $c_{t+1}=j$.  
    Applying Lemma~\ref{l:non-unique}, we get $E[c_{t+1}]=i/2\leq c_t/2$.

    In the iteration $t+1$, we either have $r_{t+1}\leq c_{t+1}$ or $r_{t+1}>c_{t+1}$. In the first case, 
    $X_{t+2}\leq X_{t+1}=c_{t+1}$. Thus $E[X_{t+2}]\leq E[c_{t+1}]\leq c_t/2\leq\max\{r_t,c_t\}/2=X_t/2$.
    If $r_{t+1}>c_{t+1}$, then $\tuple{\bm{p},\bm{e}_j}$ for some $j<r_{t+1}$ and any $\bm{p}\in\Delta_R$ by 
    Lemma~\ref{l:NE-cases}. Consequently, the row player's best response 
    $i=\pbr_r(\bm{e}_j)<j<r_{t+1}$ becomes the minimum of $R_{t+2}$, i.e., $r_{t+2}=i$.
    Applying the column analog of Lemma~\ref{l:non-unique}, we get $E[r_{t+2}]=j/2\leq r_{t+1}/2$.
    As $c_{t+2}\leq c_{t+1}$, we have $X_{t+2}\leq \max\{r_{t+2},c_{t+1}\}$.
    Consequently, $E[X_{t+2}]\leq \max\{r_{t+1}/2,c_t/2\}\leq \max\{r_t/2,c_t/2\}\leq X_t/2$.
    
    To sum up, we have $X_{t}-E[X_{t+2}]\geq X_t/2$. \cite[Corollary~17]{kotzing:krejza:tcs-19} implies 
    $E[T]\leq 2\ln n$ for the subsequence of odd iterations. Thus SDO needs at most $1+4\ln n$ iterations 
    to find the NE in expectation. 
\end{proof}

\begin{figure}[t]
    \centering
    \includegraphics[width=.15\textwidth]{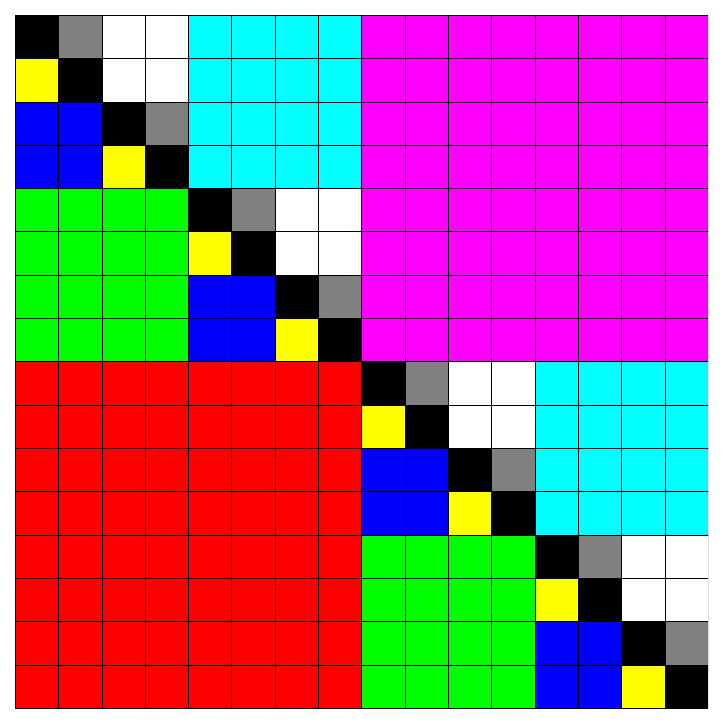}
    \includegraphics[width=.15\textwidth]{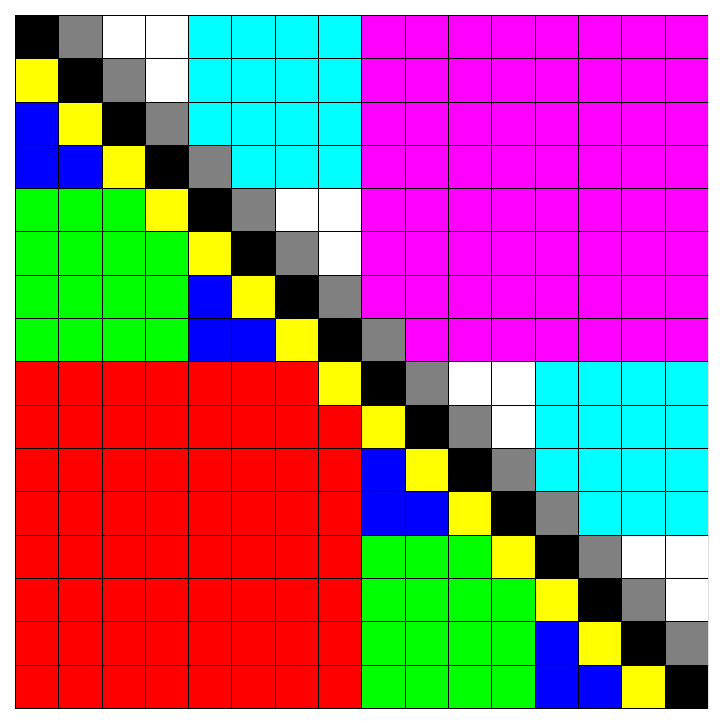}
    \caption{The clusters corresponding to the terminal states.}
    \label{fig:clusters}
\end{figure}

\begin{figure*}[t]
    \centering
    \includegraphics[width=0.3\textwidth]{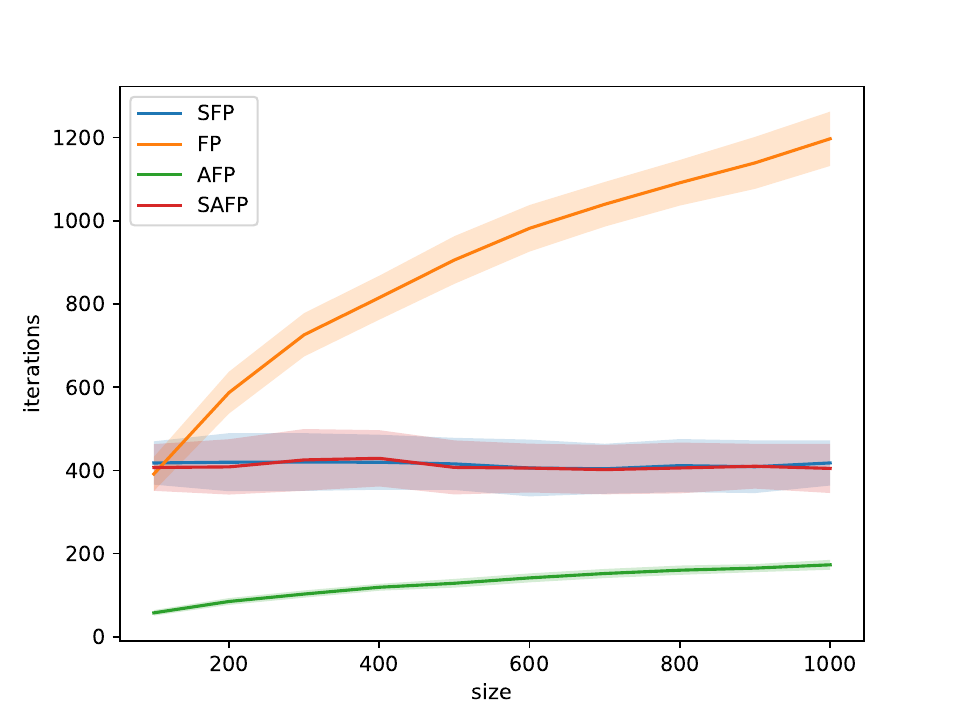}
    \includegraphics[width=0.3\textwidth]{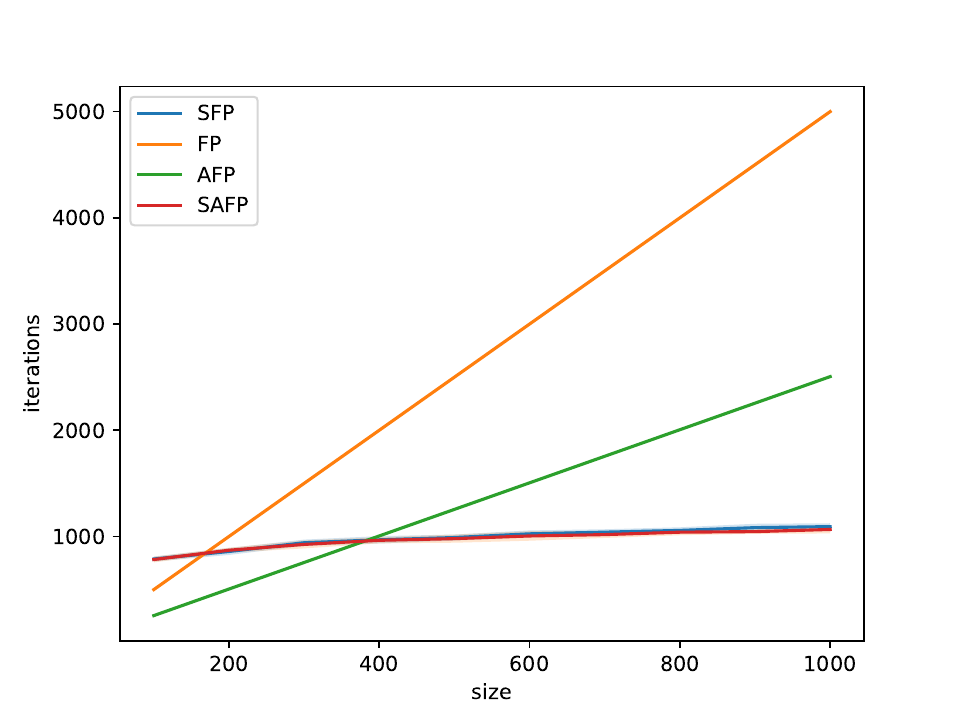}
    \includegraphics[width=0.3\textwidth]{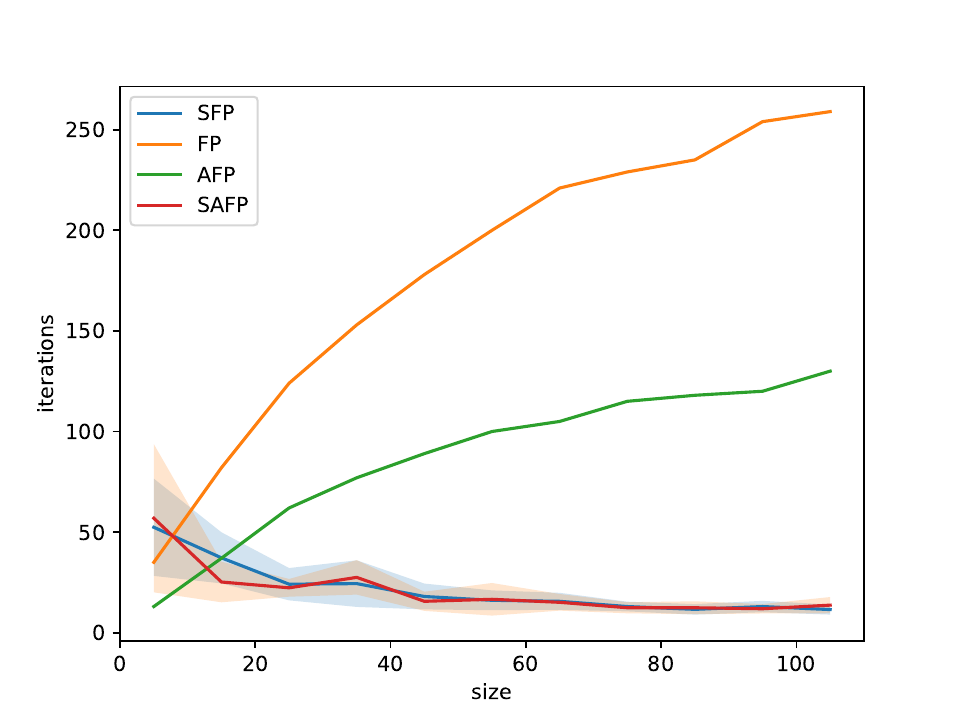}
    \caption{SFP iterations on random $[0,1]$-matrix games (left), $\bm{U}^\top$ (middle),
    and the $f$-finger Morra game (right) for $\varepsilon=0.1$.}
    \label{fig:SFP}
\end{figure*}

\begin{example}\label{ex:unique}
The second example from~\cite{zhang:sandholm:ijcai-24} is a modification of the game from 
Example~\ref{ex:non-unique} where the best responses are unique. We will again use its dual 
variant. Its matrix $\bm{U}\in\mathbb{R}^{n\times n}$ is defined as follows:
\[
\bm{U}=\left(
    \begin{array}{ccccc}
        0 & -2 & -1 & \cdots & -1\\
        2 &  \phantom{-}0 & -2 & \cdots & -1\\
        1 & \phantom{-}2 & \phantom{-}0 & \cdots & -1\\
        \vdots & \phantom{-}\vdots & \phantom{-}\vdots & \ddots & \phantom{-}\vdots\\
        1 & \phantom{-}1 & \phantom{-}1 & \cdots & \phantom{-}0
    \end{array} 
\right)
\]
Again, $\bm{U}$ has unique $\varepsilon$-NE for $0\leq\varepsilon<1$ consisting of the first row and column. 
DO needs $n$ iterations to find the NE, if initialized with $k=l=n$.
\end{example}

The following theorem can be proven analogously as Theorem~\ref{th:SDO-non-unique}. The main difference is 
a modification of Lemma~\ref{l:non-unique}; Lemma~\ref{l:unique} below (its proof is in the extended version of the paper). 
It shows that the perturbed best responses 
w.r.t. a single row $\bm{e}_k$ are not selected uniformly from $\{1,\ldots,k-1\}$ as was the case for the game 
$\bm{S}$. The distribution is due to the unique
best responses shifted towards larger indexes. However, the expected index is less than $\sfrac{3}{4}k$ which 
is sufficient to apply \cite[Corollary~17]{kotzing:krejza:tcs-19}.

\begin{theorem}\label{th:SDO-unique}
    SDO finds NE of $\bm{U}$ in $O(\log n)$ expected number of iterations. 
\end{theorem}

\begin{lemma}\label{l:unique}
    Let $\bm{x}=\tuple{1,\ldots,1,2,0,-2,-1\ldots,-1}\in\mathbb{R}^n$ be 
    the $k$-th row of $\bm{U}$, $k>1$, and 
    $\bm{z}=\tuple{z_1,\ldots,z_n}$ a random vector whose components 
    $z_i\sim\uniform(-1,1)$. 
    Let $I$ be the random variable $I=\argmax_{i\in[n]}\pi_i(\bm{x}+\bm{z})$.
    Then $E[I]\leq \sfrac{3}{4}k$.
\end{lemma}

\begin{figure*}
    \centering
    \includegraphics[width=0.3\textwidth]{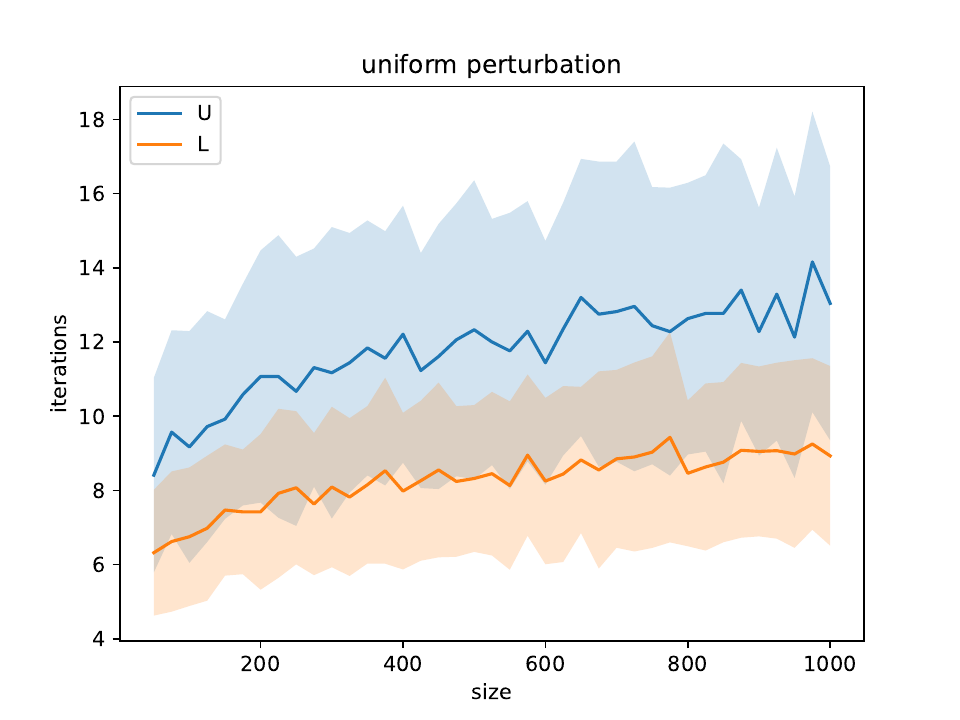}
    \includegraphics[width=0.3\textwidth]{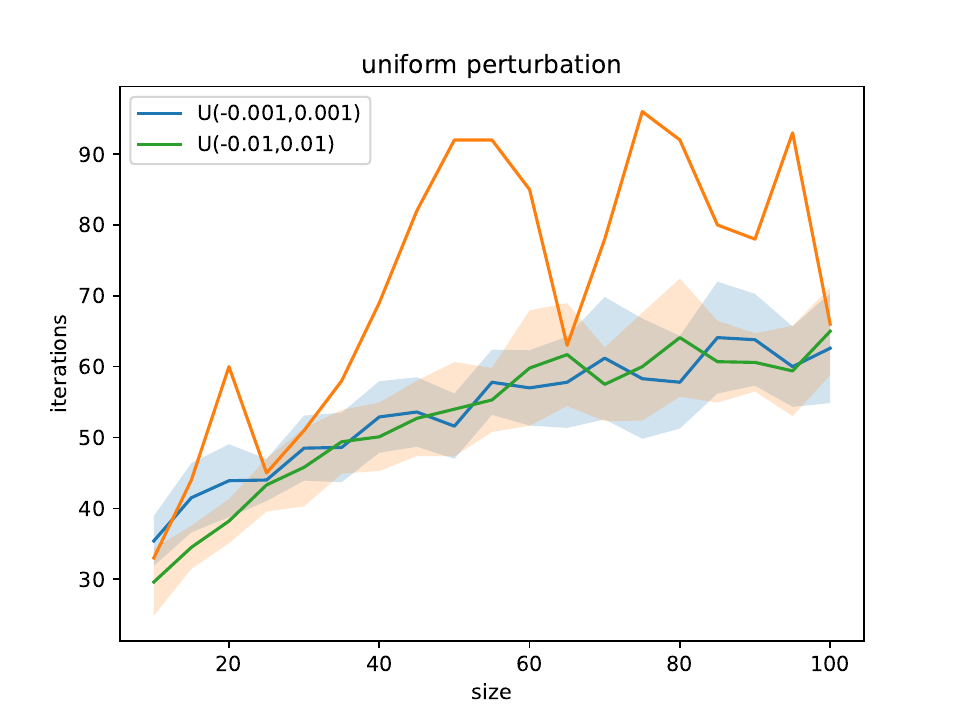}
    \includegraphics[width=0.3\textwidth]{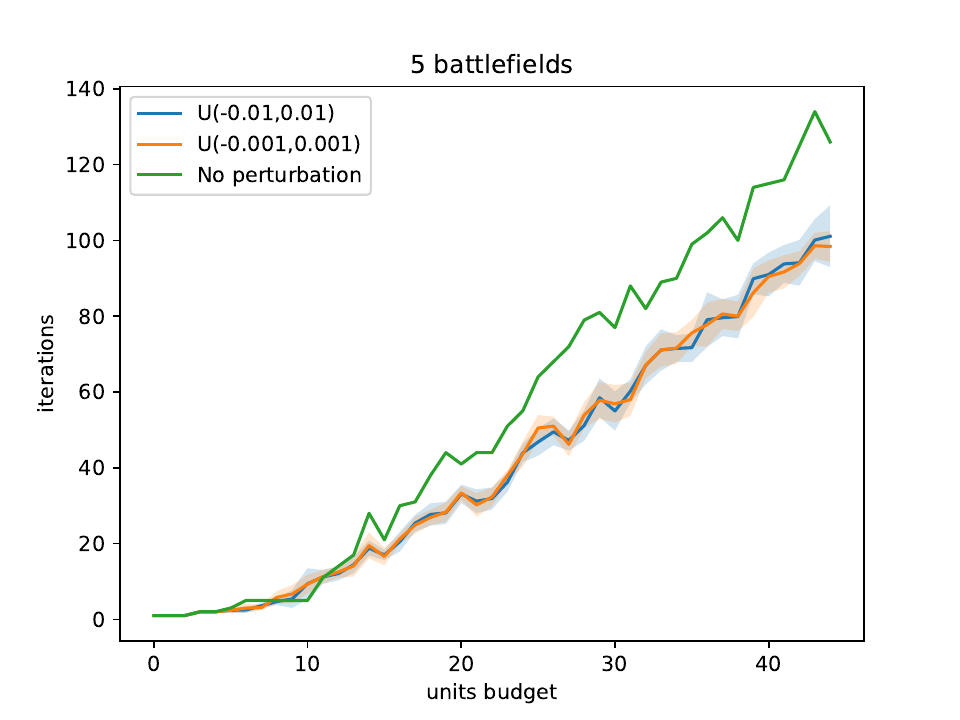}
    \caption{SDO iterations on $\bm{L}$ and $\bm{U}^\top$ with perturbations from $\uniform(-1,1)$ (left), $f$-finger Morra game (middle), and Colonel Blotto (right) for $\varepsilon=0.1$.}
    \label{fig:SDO}
\end{figure*}


\begin{figure*}[t]
    \centering
    \includegraphics[width=0.33\textwidth]{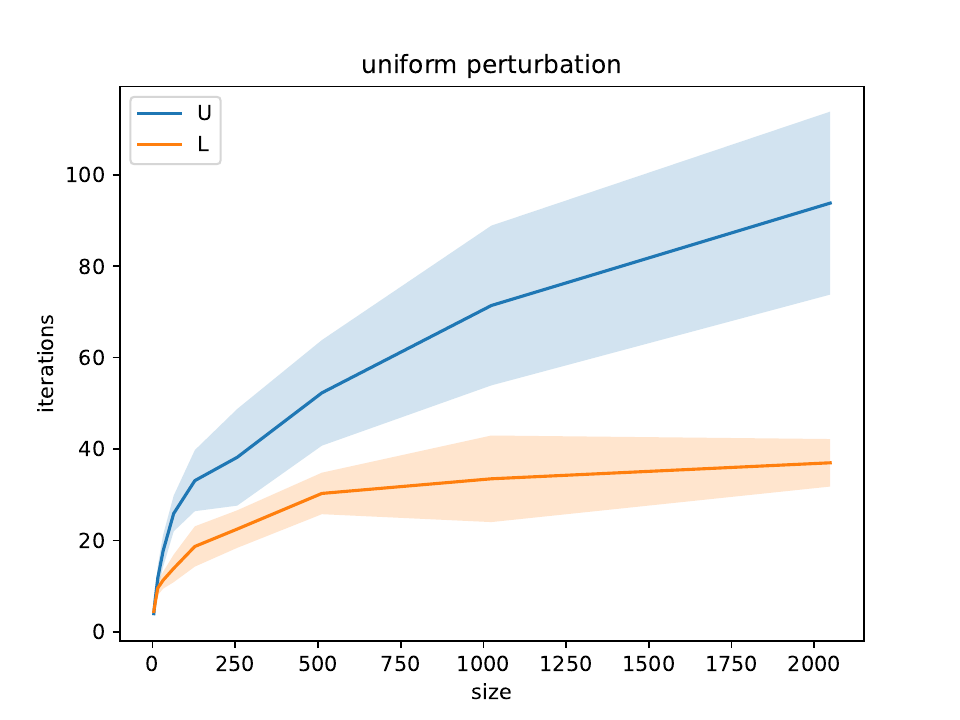}
    \resizebox{0.26\textwidth}{!}{
\begin{tikzpicture}[
    every node/.style={draw, minimum size=0.75cm},
    state/.style={circle},
    node distance=2cm
  ]
  \node[state] (a) {S};
  \node[state, right=of a] (b) {};
  \node[state, right=of b] (c) {};
  \node[state, right=of c] (d) {};

  \node[state, above=of a] (e) {};
  \node[state, above=of b] (f) {};
  \node[state, above=of c] (g) {};
  \node[state, above=of d] (h) {};

  \node[state, above=of e] (i) {};
  \node[state, above=of f] (j) {};
  \node[state, above=of g] (k) {};
  \node[state, above=of h] (l) {};

  \node[state, above=of i] (m) {};
  \node[state, above=of j] (n) {};
  \node[state, above=of k] (o) {};
  \node[state, above=of l] (p) {E};
  \draw[->, color = blue] (a) to node [draw = none, midway, xshift = 16pt, color = blue] {1/6} (e);

  \draw[->, color = blue] (a) to node [draw = none, midway, yshift = 8pt, color = blue] {1/6} (b);

  \draw[->, color = blue] (l) to node [draw = none, midway, xshift = 16pt, color = blue] {1/6} (p);

  \draw[->, color = blue] (o) to node [draw = none, midway, yshift = 8pt, color = blue] {1/6} (p);
  \draw[->, color = purple] (b) to node [draw = none, midway, xshift = 16pt, color = purple] {1/2} (f);

  \draw[->, color = purple] (b) to node [draw = none, midway, yshift = 8pt, color = purple] {1/2} (c);

  \draw[->, color = purple] (e) to node [draw = none, midway, xshift = 16pt, color = purple] {1/2} (i);

  \draw[->, color = purple] (e) to node [draw = none, midway, yshift = 8pt, color = purple] {1/2} (f);

  \draw[->, color = purple] (n) to node [draw = none, midway, yshift = 16pt, color = purple] {1/2} (o);

  \draw[->, color = purple] (k) to node [draw = none, midway, xshift = 16pt, color = purple] {1/2} (o);

  \draw[->, color = purple] (k) to node [draw = none, midway, yshift = 8pt, color = purple] {1/2} (l);

  \draw[->, color = purple] (h) to node [draw = none, midway, xshift = 16pt, color = purple] {1/2} (l);
  \draw[->] (i) to node [draw = none, midway, xshift = 16pt] {5/6} (m);

  \draw[->] (i) to node [draw = none, midway, yshift = 8pt] {5/6} (j);

  \draw[->] (f) to node [draw = none, midway, xshift = 16pt] {5/6} (j);

  \draw[->] (f) to node [draw = none, midway, yshift = 8pt] {5/6} (g);

  \draw[->] (c) to node [draw = none, midway, xshift = 16pt] {5/6} (g);

  \draw[->] (c) to node [draw = none, midway, yshift = 8pt] {5/6} (d);

  \draw[->] (j) to node [draw = none, midway, xshift = 16pt] {5/6} (n);

  \draw[->] (j) to node [draw = none, midway, yshift = 8pt] {5/6} (k);

  \draw[->] (m) to node [draw = none, midway, yshift = 8pt] {5/6} (n);
  \draw[->] (g) to node [draw = none, midway, yshift = 8pt] {5/6} (h);
  \draw[->] (g) to node [draw = none, midway, xshift = 16pt] {5/6} (k);
  \draw[->] (d) to node [draw = none, midway, xshift = 16pt] {5/6} (h);

\end{tikzpicture}}\hspace{2ex}
    \includegraphics[width=0.33\textwidth]{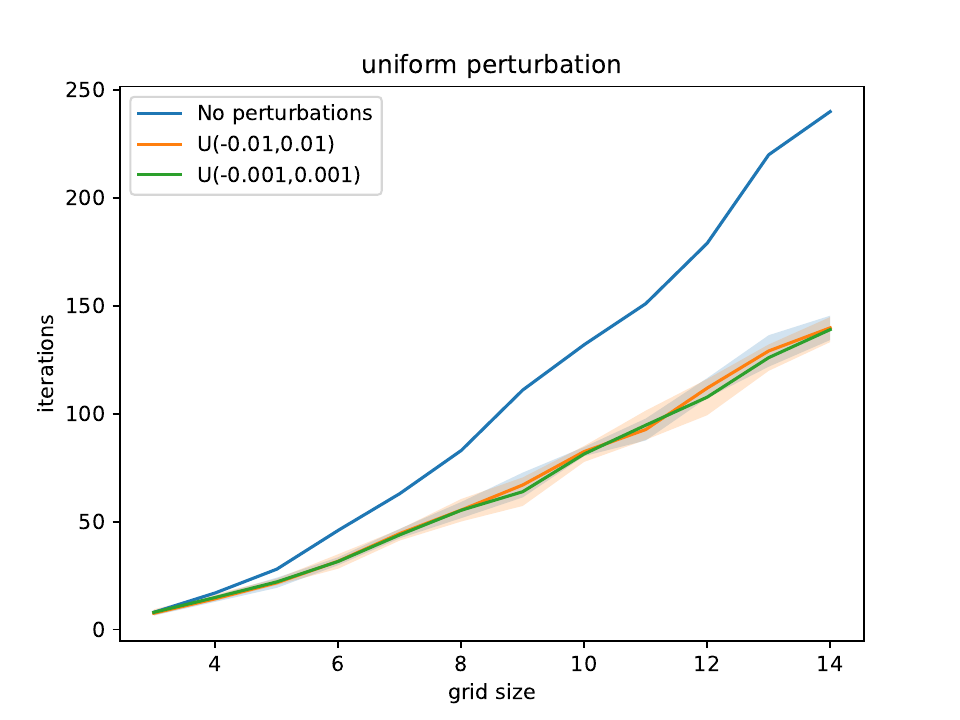}

\caption{SDO iterations for $\bm{L}$ and $\bm{U}^\top$ with efficient perturbations from $\uniform(-1,1)$ (left), $4\times 4$-grid with the transition costs for the path-planning game (middle), and the corresponding SDO iterations with uniform perturbations (right).}
\label{fig:grid}
\end{figure*}
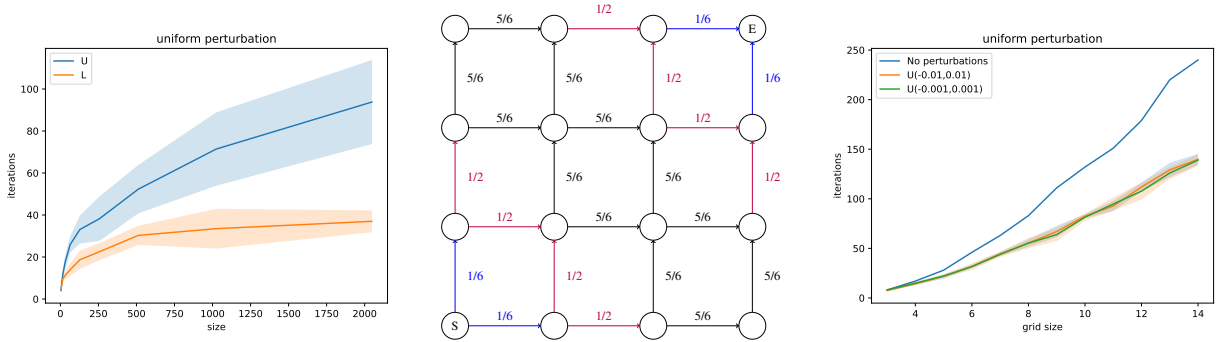

\section{Efficient Perturbations}

Let $\bm{M}\in\mathbb{R}^{m\times n}$ be a matrix game and $\bm{q}\in\Delta_n$. 
Suppose we have a best-response oracle computing $\br_r$. 
To implement a perturbed best response $\pbr_r(\bm{q})$ using $\br_r$, 
it requires perturbing every component of $\bm{M}\bm{q}$.
That cannot be done if SDO maintains only a fraction of $\bm{M}$.
Nevertheless, suppose $\bm{M}$ represents a game with inner structure like an extensive-form game (EFG) or a partially-observable stochastic game (POSG). In that case, we can implement an efficient PBRO using a best-response oracle and perturbation of utilities/rewards for terminal states or transitions.

Consider the (fully observable) stochastic game from Figure~\ref{fig:non-unique}. In this game, each player chooses a sequence 
of four bits that corresponds to the player's actions taken in non-terminal (circular) states. The terminal 
(squared) states contain their corresponding reward. This game and its generalization for any number of bits $n$
was defined in~\cite{zhang:sandholm:ijcai-24}. 
Considering the $n$-bit game, the equivalent matrix game is of size $2^n\times 2^n$ and has the form 
of the matrix $\bm{L}$ from Example~\ref{ex:non-unique}.
To perturb best responses efficiently in this game, one can perturb only the rewards of the terminal states before
applying the standard best-response oracle.

We experimentally tested that this kind of efficient perturbation speeds up the convergence of SDO
for the stochastic game from Figure~\ref{fig:non-unique}. Furthermore, we did the same for the POSG from Figure~\ref{fig:unique}
also coming from~\cite{zhang:sandholm:ijcai-24}. In this POSG, the players cannot observe the state of the game.
So they again choose only four bits. The $n$-bit generalization of this game has an equivalent matrix game 
of size $2^n\times 2^n$ and has the form of matrix $\bm{U}^\top$; see Example~\ref{ex:unique}.
The experimental results are shown in the next section. Here, we explain the details of our implementation.

One possibility to test our hypothesis for the game in Figure~\ref{fig:non-unique}
would be implementing the best response oracle as an MDP solver, for instance using the value iteration algorithm (VI).
However, VI would eliminate the existence of multiple best responses as it computes the optimal decision for 
each non-terminal state. For example, if the second player chooses $0$ as her first bit, any sequence starting 
with $1$ is a best response for the first player. However, VI would return $1,1,1,1$ as the only best response.

Thus, we tested the hypothesis on the corresponding matrix game $\bm{L}$.
Each matrix entry corresponds to a terminal state. However, this correspondence is not one-to-one, as several entries correspond to the same terminal state. So, it induces a clustering of entries in $\bm{L}$ according to the 
terminal states. The clustering for $\bm{L}$ is shown in Figure~\ref{fig:clusters} (left). Each color represents 
a single cluster.

Now, perturbing the reward of a terminal state is the same as perturbing its corresponding cluster by the same 
random value. Let $K$ be the number of clusters and $\bm{z}$ a $K$-dimensional random vector whose components 
$z_i$ are i.i.d. random variables.
For each cluster $k\in[K]$, we denote the $\{0,1\}$-matrix $\bm{B}_k\in\mathbb{R}^{m\times n}$
that masks the cluster $k$, i.e., $\bm{B}_k(i,j)=1$ iff the $i,j$-entry belongs to the cluster $k$ and 
$\bm{B}_k(i,j)=0$ otherwise. Finally, for a column player's mixed strategy $\bm{q}\in\Delta_n$, we 
compute the perturbed best-response as follows:  

\begin{equation}\label{eq:clusters}
\pbr_r(\bm{q})=\argmin_{i\in[m]} \pi_i\left(\left(\bm{L}+\sum_{k=1}^K z_k\bm{B}_k\right)\bm{q}\right)
\end{equation}

The best response for the column player is defined analogously. The same clustering can be applied also for the 
POSG in Figure~\ref{fig:unique}. The corresponding clustering is shown in Figure~\ref{fig:clusters} (right). 

\section{Experiments}

All our experiments were implemented in the programming language
Python and experimentally evaluated on the CPU AMD Ryzen 7 PRO 7840U.
All randomized algorithms were repeated ten times to report the mean and the standard
deviation. The initialization of SFP and SDO was deterministic, with the algorithm starting at the worst 
possible indexes. For tie-breaking best responses, the one with the least index was taken.
To generate pseudorandom values, we used the numpy library with the initial seed $1$.

SFP was tested in combination with Gumbel perturbations $G(0,\beta)$ where $\beta$ was set 
according to Theorem~\ref{th:SFP}. 
We ran SFP with $\varepsilon=0.1$ on random square $[0,1]$-matrix games, 
the matrix games $\bm{L}$ (Example~\ref{ex:non-unique}) and
$\bm{U}^\top$ (Example~\ref{ex:unique}), and $f$-finger Morra game \cite{good:siam-65} all normalized to the interval $[0,1]$;
Figure~\ref{fig:SFP}. We compared SFP with FP, AFP, and AFP with perturbed best responses (SAFP). 
The left graph shows the number of iterations on random $n\times n$-matrix games.
In the experiment, we generated 100 random games for each $n$. AFP outperforms all the other methods; 
SFP and SAFP behave similarly on random games.   
The middle graph presents the number of iterations for the game $\bm{U}^\top$. An analogous graph for 
$\bm{L}$ looks almost identical, so we omit it. FP and AFP show their the worst-case $O(n)$ complexity. 
SFP and SAFP again have a similar performance. 
Finally, the right graph shows the number of iterations for the $f$-finger Morra game.
The horizontal axis represents the number of fingers $f$. The size of the corresponding matrix is $f^2\times f^2$. 
Interestingly, SFP and SAFP are able to quickly find an $\varepsilon$-NE for large values of $f$. 
However, this is not the case for FP and AFP.

We did similar experiments for DO and SDO combined with the uniform perturbations. 
The SDO iterations for matrix games $\bm{L}$, $\bm{U}^\top$, $f$-finger Morra game, 
and the Colonel Blotto game with five battlefields for different
numbers of units are shown in Figure~\ref{fig:SDO}. We omit the DO iterations in the left graph as the DO needs $n$ iterations
for the size $n$. In all these cases, SDO outperforms DO. However, we noticed
this is not the case for random games where perturbations do not provide a faster convergence or even degrades the convergence 
rate if the perturbations are too large. 
This outcome is somewhat expected, considering that the best responses in random games are inherently stochastic.

Further, we tested the efficient perturbations of clusters, see Equation~\eqref{eq:clusters}, 
on the matrix games $\bm{L}$ and $\bm{U}^\top$ of size $2^n\times 2^n$
corresponding to stochastic games from Figure~\ref{fig:non-unique} and \ref{fig:unique}.
The results are shown in Figure~\ref{fig:grid} (left). Although SDO with efficient perturbations
requires more iterations in comparison with Figure~\ref{fig:SDO} (left), still the obtained results are 
much better than the linear complexity of DO.

To test the efficient perturbations further, we defined a path-planning game on an $n\times n$-grid.
An example of such a grid for $n=4$ is shown in Figure~\ref{fig:grid} (middle). The colors and numbers 
denote the costs for particular transitions. The path-planning player looks for the shortest path 
in the grid starting in the node S and finishing in the node E. The other player selects a single edge
and multiplies its cost by a given coefficient; $10$ in our experiments. We implemented the 
perturbed best-response oracle by perturbing the transition costs every time the oracle is called.
Figure~\ref{fig:grid} (right) show that uniform perturbations reduce the number of 
iterations with increasing grid size.


\section{Conclusions}

To summarize, perturbing best responses improves the convergence of FP and DO.
We proved that SFP has a logarithmic complexity in the number of pure strategies $n$.
Although DO needs $\Theta(n)$ many iterations to find $\varepsilon$-NE in the worst case, 
it is still possible that SDO finds an $\varepsilon$-NE in $O(\log n)$ expected number of iterations.
We leave this open problem for future research.

\newpage
\section{Acknowledgments}
The authors were supported by the Czech Science Foundation grant--no. 24-12046S.


\newpage

\appendix

\section{Appendix}

\setcounter{theorem}{7}
\begin{lemma}
    Let $\bm{x}=\tuple{1,\ldots,1,2,0,-2,-1\ldots,-1}\in\mathbb{R}^n$ be 
    the $k$-th row of $\bm{U}$, $k>1$, and 
    $\bm{z}=\tuple{z_1,\ldots,z_n}$ a random vector whose components 
    $z_i\sim\uniform(-1,1)$. 
    Let $I$ be that random variable $I=\argmax_{i\in[n]}\pi_i(\bm{x}+\bm{z})$.
    Then $E[I]\leq \sfrac{3}{4}k$.
\end{lemma}
\begin{proof}
    We have $E[I] = \sum_{i=1}^n i\cdot P(I=i)=\sum_{i=1}^{k-1} i\cdot P(I=i)$ as $P(I=i)=0$ for 
    $i\geq k$ by the choice of $\delta$. 

    We first compute $P(I=k-1)$. We have $I=k-1$ iff $2+z_{k-1}>1+z_i$ for all 
    $i<k-1$ iff $1+z_{k-1}>z_i$ for all 
    $i<k-1$. As $z_i$ are drawn i.i.d., the conditional probability 
    $P(I=k-1|z_{k-1})=\left[F(1+z_{k-1})\right]^{k-2}$ where 
    $F(x)$ is the CDF for $\uniform(-1,1)$, i.e., $F(x) = \sfrac{1}{2}(1+x)$ for 
    $x\in [-1,1]$ and $F(x)=1$ for $x\geq 1$. The corresponding PDF is denoted $f$, recall that 
    $f(x)=\sfrac{1}{2}$ for $x\in[-1,1]$. To get rid of 
    the conditioning on $z_{k-1}$, we integrate
    \begin{align*}
        P(I=k-1) &= \int_{-1}^{1}f(z_{k-1})
                    \left[F(1+z_{k-1})\right]^{k-2}dz_{k-1}\\
                 &= \sfrac{1}{2}\int_{-1}^{1}
                    \left[F(1+z_{k-1})\right]^{k-2}dz_{k-1}\\
                 &= \sfrac{1}{2}+\sfrac{1}{2}\int_{-1}^{0}
                    \left[F(1+z_{k-1})\right]^{k-2}dz_{k-1}\\
                 &= \sfrac{1}{2}+(\sfrac{1}{2})^{k-1}\int_{-1}^{0}
                    \left[2+z_{k-1}\right]^{k-2}dz_{k-1}\\
                 &= \sfrac{1}{2}+(\sfrac{1}{2})^{k-1}\int_{1}^{2}
                    y^{k-2}dy\\
                 &= \sfrac{1}{2}+(\sfrac{1}{2})^{k-1}\left[\frac{y^{k-1}}{k-1}\right]_{1}^{2}\\
                 &= \sfrac{1}{2}+\frac{1-(\sfrac{1}{2})^{k-1}}{k-1} \leq \sfrac{1}{2}+\frac{1}{k-1}
    \end{align*}

    Next, we compute $P(I=j)$ for $j<k-1$. $I=j$ iff $z_{j}>z_i$ 
    for all $j\neq i<k-1$ and $-1+z_{j}>z_{k-1}$. 
    As $z_i$ are drawn i.i.d., the conditional probability 
    $P(I=j|z_{j})=F(z_j)^{k-3}\cdot F(-1+z_{j})$.
    Thus, we have
    \begin{align*}
        P(I=j) &= \int_{-1}^{1}f(z_{j})
                    F(z_j)^{k-3}F(-1+z_{j})dz_{j}\\
                 &= (\sfrac{1}{2})^{k-1}\int_{-1}^{1}
                    (1+z_{j})^{k-3}z_j dz_{j}\\
                 &= (\sfrac{1}{2})^{k-1}\int_{0}^{2}
                    y^{k-3}(y-1)dy\\
                 &= (\sfrac{1}{2})^{k-1}\left(\int_{0}^{2}
                    y^{k-2}dy - \int_{0}^{2}y^{k-3}dy\right)\\
                 &= (\sfrac{1}{2})^{k-1}\left(\left[\frac{y^{k-1}}{k-1}\right]_0^2 - 
                 \left[\frac{y^{k-2}}{k-2}\right]_{0}^{2}\right)\\
                 &= \frac{1}{k-1}-\frac{1}{2(k-2)}=\frac{k-3}{2(k-1)(k-2)}\\
    \end{align*}

    Combining the above computations, we have
    \begin{align*}
        E[I] &\leq \frac{k-3}{2(k-1)(k-2)}\sum_{i=1}^{k-2}i + 
              (k-1)\cdot\left(\sfrac{1}{2}+\frac{1}{k-1}\right)\\
             &= \frac{k-3}{2(k-1)(k-2)}\frac{(k-1)(k-2)}{2} + 
              \sfrac{1}{2}(k-1)+1\\
             &= \frac{k-3}{4} + 
              \sfrac{1}{2}k+\sfrac{1}{2}=\sfrac{3}{4}k-\sfrac{1}{4}\leq\sfrac{3}{4}k\\
    \end{align*}
\end{proof}

\end{document}